\documentstyle[aps,psfig,prb,floats]{revtex}

\ifx\undefined\psfig\def\psfig#1{ }\else\fi

\begin{document}
\ifpreprintsty\else
\twocolumn[\hsize\textwidth%
\columnwidth\hsize\csname@twocolumnfalse\endcsname
\fi

\title{Field theory of Skyrme lattices 
in quantum Hall ferromagnets}
\author{M. Abolfath$^{1,2}$, and M. R. Ejtehadi$^2$}
\address{Center for Theoretical Physics and Mathematics, 
P.O.Box 11365-8486, Tehran, Iran}
\address{
Institute for Studies in Theoretical Physics and Mathematics
P.O. Box 19395-5531, Tehran, Iran
}
\date{\today}
\maketitle

\begin{abstract}
\leftskip 2cm
\rightskip 2cm

We report the application of the nonlinear $\sigma$ model
to study the multi-skyrmion problem in the
quantum Hall ferromagnet system. 
We make use of a first-principle calculation to
derive an analytical form for the inter-skyrmionic interaction to
show that the ground
state of the system can be described by a 
ferromagnet triangular Skyrme lattice
near $\nu=1$ where skyrmions are extremely dilute 
and a continuous transition into 
antiferromagnet square lattice occurs by increasing the 
skyrmion density and therefore $|\nu-1|$.
Using these results we demonstrate that the transition for a triangular to
a square lattice which was previously derived, using the Hartree-Fock
method, can also be seen in the field theory picture.
We investigate the possibility that the skyrmions bound in pair 
to make a bi-skyrmion triangular lattice when the Zeeman energy
is extremely small.
We show that the energy of a skyrmion with charge $Q$ is less than the 
energy of $Q$ skyrmions each with charge one when 
the short range interaction among them is considered.
By taking the quantum fluctuations into account,
we also argue the possibility of the existence of a 
superconductor-insulator and the
non-zero temperature phase transitions. 
\end{abstract}


\ifpreprintsty\else\vskip1pc]\fi
\narrowtext

\section{Introduction}
Recently, there have been a number of experimental 
\cite{barrett,jpe-skyrme,optical1,bayot}
and theoretical \cite{Sondhi,Moon,Yang,Fertig94,Abolfath} 
investigations of skyrmions in the integer and 
fractional quantum Hall effect (QHE) in 
2-dimensional electron gas (2DEG). It has been reported \cite{Brey,Cote97}
that the microscopic calculations show
the ground state of a two-dimensional electron system at
the Landau level filling factor near $\nu=1$ is a Skyrme (soliton) lattice. 
Recently Brey {\em et al.} \cite{Brey} and C\^ot\'e {\em et al.}
\cite{Cote97} have made use of the time-dependent
random phase approximation (RPA)
to show that the zero-temperature ground state of a 
multi-skyrmion system is a centered square lattice. In particular
they have found the possibility of the structural phase transition
into triangular Wigner crystal by changing the Zeeman energy and the
Landau level filling factor.
One may expect on the general grounds, 
these microscopic observations should be
supported by other many body techniques, e.g., the appropriate field theory.
It has been shown by
Sondhi {\em et al.} \cite{Sondhi} and Moon {\em et al.} 
\cite{Moon}, that
the quantum Hall ferromagnet can be described by a generalized minimal
nonlinear $\sigma$ (NL$\sigma$) model to add the Zeeman and Coulomb energy,
as an alternative approach. The consistency of this minimal field theory 
with other microscopic calculations for a system of a
single skyrmion has been reported numerically
by Abolfath {\em et al.}. \cite{Abolfath}
Recently, Green {\em et al.} \cite{Green} have exploited a variational
approach on the generalized NL$\sigma$-model to study the multi-skyrmion
system. They have indicated that the
ground state is closer to a hexagonal structure, i.e.,
a distorted centered square lattice.
Surprisingly, they came to a quite different conclusion from those
obtained by microscopic consideration. 
One naturally seeks the reasons for the reported discrepancies between the 
different approaches employed for the multi-skyrmion systems.
The aim of this paper is to present a proper field theory picture which
avoids such controvertial issues.
By exploiting the minimal field theory of the
quantum Hall NL$\sigma$-model \cite{Sondhi,Moon,Abolfath}
we present a description for the field theory which is 
consistent with the microscopic picture.
In this way, we present a {\em first-principle} calculation
to derive the analytical form for the inter-skyrmionic interaction.
To the best of our knowledge this type of calculation has not been reported
before. In the past, several authors have 
used a phenomenological approach to guess the form
of the interaction by using the known Hartree-Fock long wave length
spin waves spectrum. \cite{Shankar,Cote97,Nazarov}
In contrast to Green {\em et al.} \cite{Green}, 
we explicitly show that the ground state
is a centered square lattice and/or triangular Wigner crystal,
deponding on the value of the Zeeman energy 
and/or Landau level filling factor. 
These results which were previously derived, using the Hartree-Fock
method,\cite{Brey,Cote97} can also be seen in our field theory picture.
We start from a proper superposition rule for the unit vectors
of the NL$\sigma$-model to evaluate the correct analytical form
for the inter-skyrmionic interaction.
We therefore demonstrate that the field theoretic description
of the NL$\sigma$-model does indeed lead to conclusion in agreement
with microscopic Hartree-Fock calculations.
We take advantage of the particle-hole symmetry
\cite{Fertig94} (which is equivalent 
to skyrmion-antiskyrmion duality 
in the classical field theory) 
to study the skyrmions associated with holes ($\nu \leq 1$).
We will show that within $\nu_{1c} < \nu \sim 1$, 
the ground state is a triangular lattice  
in order to reduce the Coulomb energy, since
the effective interaction energy between skyrmions through 
the channel of the Zeeman term is negligible. 
Increasing the skyrmion 
density in such a way that $0 << \nu_{2c} < \nu < \nu_{1c}$, 
leads to the formation of 
a centered square lattice. 
The square lattice is favored by the effective interaction made by
the gradient and Zeeman energy.
The effective interaction which is the attractive XY model
allows the antiferromagnetic 
alignment of the skyrmion's orientation.
This can be anticipated by the frustration of the
skyrmionic hedge-hog fields within the triangular 
lattice which is the 
favorable orientation for the Coulomb energy.
This leads to a structural phase transition due to 
varying 
the Landau level filling factor.
The attractive XY interaction makes a tendency for collapsing the 
single skyrmions which are located in different positions. 
It favors recombination of the $N$-seperated single skyrmions 
and forming a single skyrmion with topological charge $N$.
The competition between the attractive XY interaction and the repulsive 
Coulomb interaction leads to stability of the Skyrme lattice of
the single-skyrmions.
We will discuss for any short range interaction, like the hard-core model,
the strength of the repulsive interaction is not sufficient to make
a stable lattice. In this case, the energy of a charge-Q skyrmion is
lower than any other configuration, hence the destruction of the lattice.
We demonstrate in special situations, at
extremely small values of the Zeeman energy, 
(depends on the filling factor) a single skyrmion square lattice
(with antiferromagnets orientation) collapses to a triangular lattice
of bi-skyrmions (charge-2 skyrmions). In the latter, the orientation
of skyrmions is governed by the ferromagnets ordering.
By taking the quantum fluctuations into account, 
we also argue the possibility of the existence of a 
superconductor-insulator and a non-zero temperature
Kosterlitz-Thouless phase transition, $T_{KT}$. 
We also show that the 2DES with the typical Zeeman energy at filling factor
smaller than one (the validity range of the square lattice) 
are in the insulating phase.
We finally estimate the melting temperature, $T_m$, for the lattice
via the Kosterlitz-Thouless-Halperin-Nelson-Young (KTHNY) mechanism.
A comparison between $T_{KT}$ and $T_m$ shows that both mechanisms
can be accounted for these systems, depending on the Zeeman splitting
factor and/or the Landau level filling factor.

\section{An overview on NL$\sigma$-model in QHF}
In the NL$\sigma$-model approach to 
QHF, the spin of an electron may
be described classically by a unit vector (spin coherent state), 
as an order parameter, whose direction may be
changed continuously 
in the space. In this representation, the effective
Hamiltonian is functional of the unit vector ${\bf m}({\bf r})$
\cite{Sondhi,Moon}
\begin{equation}
E[{\bf m}]=E_0[{\bf m}]+E_z[{\bf m}]+E_c[{\bf m}],
\label{eq1}
\end{equation}
where $E_0[{\bf m}]$ and $E_z[{\bf m}]$ are the conventional NL$\sigma$ model 
and the Zeeman energy respectively. 
The last term $E_c[{\bf m}]$ is the long range Coulomb energy 
due to the equivalence between Hall current 
and skyrmion topological current\cite{Sondhi,Moon}
\begin{mathletters}
\label{eq2}
\begin{equation}
E_0[{\bf m}]=\frac {\rho_s}{2}\; \int d^{2}r\, (\nabla{\bf m})^2, 
\label{eq2a}
\end{equation}
\begin{equation}
E_z[{\bf m}]=\frac{\tilde{g}}{2 \pi \ell_0^2} 
\int d^{2}r\, [ \; 1 - m_z({\bf r}) \; ], 
\label{eq2b}
\end{equation}
\begin{equation}
E_c[{\bf m}]={e^2\over 2 \epsilon} \int d^{2}r \int d^{2}r^\prime \,
\frac{\rho({\bf r})\rho({\bf r}^\prime)}{|{\bf r}-{\bf r}^\prime|}.
\label{eq2c}
\end{equation}
\end{mathletters}
The long range nature of the Coulomb interaction is crucial for stability
of the Skyrme lattice. We will get back to this point in Sec. \ref{Recomb}.
Here $\rho_s=e^2/(16\sqrt{2\pi}\epsilon\ell_0)$ is the spin stiffness
at $\nu=1$
(assuming zero layer thickness for the 2DEG) which arises from the Coulomb
exchange energy,
$\epsilon$ is the background dielectric constant of the semiconductor,
$\tilde{g}=g e^2/(2\epsilon\ell_0)$ is the Zeeman term, $g$ is the
effective gyromagnetic ratio, and $\ell_0$ is the magnetic length. 
The charge density is given by \cite{Sondhi,Moon}
$\rho({\bf r})=\frac{-\nu}{8 \pi} \; \epsilon_{\alpha \beta} \; 
{\bf m({\bf r})} 
\cdot [\partial_\alpha{\bf m({\bf r})} \times 
\partial_\beta{\bf m({\bf r})}]$ 
which is equal to the filling factor times
the topological $O(3)$ spin texture density of the 
quantum Hall ferromagnet (QHF). 
The total skyrmion charge denoted by $Q$
is an integer-valued topological invariant. It can be determined
by the integration upon the charge density and classified by homotopy group
of a 2D-sphere respectively.
The lowest energy skyrmion solution, ${\bf \tilde{m}}({\bf r})$,
has to satisfy a non-linear differential equation which can be obtained
by minimizing the energy in Eq.(\ref{eq1}) with respect to ${\bf m}$
\begin{eqnarray}
&\rho_s& \Bigl( -\nabla^2  + {\bf \tilde{m}} \cdot 
\nabla^2 {\bf \tilde{m}} \Bigl){\rm \tilde{m}}_\mu 
-\frac{\tilde{g}}{2\pi \ell_0^2} (\delta_{z\mu}-
{\rm \tilde{m}}_z{\rm \tilde{m}}_\mu) \nonumber \\&&
- \frac{\nu}{4\pi} \epsilon_{\alpha \beta} \{ \partial_\alpha V({\bf r}) \}
({\bf \tilde{m}} \times \partial_\beta {\bf \tilde{m}})_\mu=0,
\label{eq3}
\end{eqnarray}
where $V({\bf r})$ is the Hartree potential 
(the exchange potential resides in $\rho_s$)
\begin{equation}
V({\bf r})=\frac{e^2}{\epsilon}
\int d^2{\bf r}^\prime \frac{\tilde{\rho}({\bf r}^\prime)}
{|{\bf r}-{\bf r}^\prime|},
\label{eq4}
\end{equation}
and $\tilde{\rho}$ is the skyrmion charge density associated with the 
minimum energy solution, ${\bf \tilde{m}}({\bf r})$.
The solutions of Eq.(\ref{eq3}) can be classified by the skyrmion charge
$Q=\int d^2{\bf r} \; \rho({\bf r})$. From now 
and for the sake of simplicity, we remove 
the tilde over the classical solution and denote it by ${\bf m}$.
It is easy to find the following equation of motion of the optimal skyrmion
by making use of the cross product of ${\bf m}$ upon Eq.(\ref{eq3})
\begin{equation}
\partial_\alpha J_\alpha^\lambda = \frac{\tilde{g}}{2\pi\ell_0^2}
(\hat{z}\times{\bf m})_\lambda,
\label{eq5}
\end{equation}
where
\begin{equation}
J_\alpha^\lambda=\rho_s({\bf m}\times\partial_\alpha{\bf m})_\lambda
-\frac{\nu}{4\pi} V({\bf r}) 
\epsilon_{\alpha \beta}\partial_\beta m_\lambda.
\label{eq6}
\end{equation}
One may immediately read off from Eq.(\ref{eq5}) that 
$\partial_\alpha J_\alpha^3 = 0$, i.e., a conserved iso-spin current.
We may define the ground state of QHF at 
$\nu=1$ as a vacuum of skyrmionic spin textures 
 where all spins are aligned
along the magnetic field direction, i.e., the $\hat{z}$-axis.  
Note that in the
absence of the Zeeman energy, the alignment of spins along an arbitrary
axis occurs due to spontaneous 
global $O(3)$ symmetry breaking \cite{Moon}
hence the minimizing the electrons exchange energy.
This is also the state of spins far from the center of the skyrmions.
Note that the number of 
skyrmions (antiskyrmions) in the ground state is counted by $|N-N_\phi|$.
Before we consider a lattice of skyrmions we 
need to have the correct shape of a single skyrmion. 
A single skyrmion is a topological optimal solution of 
Eq.(\ref{eq3}) with $Q=1$.
For our purposes, it is convenient 
to parameterize the unit vector ${\bf m}$ by
\begin{equation}
{\bf m} = (\varphi_x, \varphi_y, \sqrt{1-\overline{\psi}\psi}),
\label{eq6.1}
\end{equation}
where $\psi = \varphi_x + i \varphi_y$. 
Near the core of the skyrmion we do not expect that the shape of
the skyrmion is influenced much by the magnetic field. However,
for large distances from the core, the Zeeman energy becomes
dominant and we need to consider its effect where
the direction of the unit vector ${\bf m}$ 
is close to the vacuum, namely, $\hat{z}$-axis. 
Taking the limit of small $\psi$, we can expand the ${\bf m}$ up to 
quadratic order in the $\psi$. Putting this in Eq.(\ref{eq6.1}) gives
\begin{equation}
E[\psi] = \int d^2{\bf r} \left(\frac{-\rho_s}{2} \overline{\psi}
\nabla^2 \psi + \frac{\tilde{g}}{4\pi\ell_0^2} \overline{\psi} \psi\right),
\label{eq6.2}
\end{equation}
which leads to the equation of motion
\begin{equation}
-\rho_s\nabla^2 \psi
+ \frac{\tilde{g}}{2\pi\ell_0^2} \psi = 0.
\label{eq6.21}
\end{equation}
Introducing $\kappa^2 = \tilde{g}/(2\pi\ell_0^2\rho_s)$
we simply have the equation $-\nabla^2 \psi + \kappa^2 \psi=0$.
We are interested in the `vortex' solution 
$\psi = 2\partial_{z} \Upsilon$, 
where $\partial_{z} = (\partial_x + i \partial_y)/2$ 
and $z=x+iy$. (If $\tilde{g}=0$, this would result in $\nabla^2 \Upsilon=0$ with
the solution $\Upsilon \propto \ln(r)$ and therefore $\psi \propto z/r^2$).
Substituting this we find $-\nabla^2 \Upsilon + \kappa^2 \Upsilon =0$ or
$\Upsilon \propto e^{-\kappa r}/\sqrt{r}$ and therefore that 
$\psi \propto z e^{-\kappa r}/r^{3/2}$ far from the core.
The most important of this part is of course the exponential
(as opposed to, algebraic if $\tilde{g}=0$), fall-off.\cite{Abolfath,Henk}
To check this, we solve Eq.(\ref{eq3}) for a single-skyrmion.
\cite{Abolfath} The results obtained from these numerical calculation
shows $\theta(r) = |\psi(r)| = c\kappa K_1(\kappa r)/2\pi$
far from the core of skyrmion where
$K_1(x) = \sqrt{\pi/(2x)} e^{-x} (1 + {\cal O}(1/x))$
is the modified bessel function. $c$ is a constant, can be obtained from
the asymptotic form of $\theta(r)$
\begin{equation}
c = 30.4.
\end{equation}
The dynamics of a skyrmion spin texture in NL$\sigma$-model 
may be incorporated via the Wess-Zumino action.\cite{fradkin}
The result of expansion for the single skyrmion's Wess-Zumino term is
\begin{eqnarray}
S_{WZ} &=& \frac{\hbar}{2} \int_0^{\hbar\beta} d\tau
\int dt \; {\bf m}\cdot(\partial_\tau{\bf m}\times\partial_t{\bf m})
\nonumber\\&& = \frac{\hbar}{8\pi\ell_0^2} \int_0^{\hbar\beta} d\tau
\int d^2{\bf r} \; \overline{\psi}({\bf r}, \tau)
\frac{\partial}{\partial\tau} \psi({\bf r}, \tau),
\label{12.1}
\end{eqnarray}
where we keep the quadratic terms in Eq.(\ref{12.1}). 
At this level of approximation,
the effective action may be obtained by Eq.(\ref{eq6.2}) and Eq.(\ref{12.1})
where the Lagrangian density is 
\begin{equation}
{\cal L} = \overline{\psi}({\bf r}, \tau) \left(\frac{\hbar}{8\pi\ell_0^2} 
\frac{\partial}{\partial\tau} +
\frac{\rho_s}{2} \nabla^2 - \frac{\tilde{g}}{4\pi\ell_0^2} \right)
\psi({\bf r}, \tau),
\label{eq6.3}
\end{equation}
and $S = \int d\tau \int d^2{\bf r} {\cal L}$.
The optimal solution of $\psi$ is then identical to the solution of the 
time dependent Schr\"odinger equation where the external potential is 
proportional to the Zeeman splitting factor. Then the single skyrmion behaves
approximately like a quantum mechanical point particle far from its core. 
Eq.(\ref{eq6.3}) describes the usual gapful 
ferromagnetic spin wave mode.

\section{Skyrmion interaction}

We can now consider the interaction between skyrmions by generalizing 
our linearized energy functional. For that one should note
that for the optimal single skyrmion spin texture in Eq.\ (\ref{eq3}) 
we may choose a particular orientation. 
In general the texture can be rotated without costing any 
energy. Since the system shows the $U(1)$ symmetry, any valid 
skyrmion spin texture can be obtained 
by rotating all spins about
the $\hat{z}$-axis by angle $\chi$. Therefore the state,
${\bf m}' = \exp(i \chi \hat{I}_z){\bf m}$, is also an optimal solution
of the Hamiltonian where $\hat{I}_z \equiv -i \partial/\partial\chi$ is the 
generator of the rotation along the $\hat{z}$-axis in the internal space. 
One may expect that it contributes to the Hamiltonian through 
$(\hbar^2/2\Lambda_0)(-i\partial/\partial\chi - \xi_0)^2$ where $\Lambda_0$
is the moment of inertia 
of the single skyrmion. 
This is the leading term of the total energy which is expanded 
with respect to the number of reversed spin, $\xi$.
In general the dimensional analysis shows $E_z \propto \xi$ and 
$E_c \propto 1/\sqrt{\xi}$.
The optimal value of the number of reversed spin, 
$\xi_0 = E_z/(2\tilde{g})$, can be evaluated by minimization of the
total energy with respect to $\xi$
\begin{equation}
\xi_0 \equiv \frac{1}{4\pi\ell_0^2} \int d^2{\bf r} [1-m_z({\bf r})],
\label{xi}
\end{equation}
where $m_z({\bf r})$ is the optimal solution of Eq.(\ref{eq3}) 
corresponding to the given Zeeman splitting factor. 
This leads to the optimal Coulomb and Zeeman energy and then the predicition 
of $E_c/E_z = 2$ and hence 
$\Lambda_0 \equiv \hbar^2\left(d^2 E/d\xi^2\right)^{-1}_{\xi_0}
= \hbar^2 E_z/(6\tilde{g}^2)$. The validity of this simple dimensional
analysis has been demonstrated numericaly in Ref. 9. 
Note that $\xi$ is a quantized variable since $\chi$ is compact.

\subsection{Square lattice}
\label{SubA}
Our goal is now to calculate the interaction
energy between skyrmions with different orientations. 
Following Piette {\em et al.} \cite{Piette}, 
we start with the conventional NL$\sigma$ 
model to find out the proper superposition rule for 
the hedge-hog fields of different skyrmions. 
In the absence of the Zeeman and Coulomb energies,
the energy functional is scale invariant and one may find the optimal
solutions analytically. In this case, it can be shown that any 
analytic complex polynomial defines an optimal solution 
\cite{rajaraman,Belavin,Woo} 
\begin{equation}
{\bf m(r)}=\left(\frac{2w_x}{1+|w|^2}, \frac{2w_y}{1+|w|^2},
\frac{1-|w|^2}{1+|w|^2}\right),
\label{1}
\end{equation}
where 
$w = w_x + i w_y$ is any $Q$-sector analytical function.
One may decompose $w$ into a series of analytical functions each
with $Q=1$
\begin{equation}
w = \sum_{j=1}^N u_j.
\label{2}
\end{equation}
Any $Q=1$ analytical function represent a single-skyrmion, 
then the number of skyrmions $N$ is clearly
the total skyrmions topological charge, i.e., $N=Q$.
Eq.(\ref{2}) denotes a sequence of order parameters 
${\bf m}_j({\bf r})$ in configuration space. 
It is convenient to parameterize the ${\bf m}_j({\bf r})$ by
\begin{equation}
{\bf m}_j({\bf r}) = 
\left( \sin\eta_j({\bf r})  \cos\zeta_j({\bf r}), 
\sin\eta_j({\bf r})  \sin\zeta_j({\bf r}),
\cos\eta_j({\bf r}) \right),
\label{2.1}
\end{equation}
where $\eta_j({\bf r})$ and $\zeta_j({\bf r})$ are polar and 
azimuthal field variable associated with the $j$th skyrmion.
One may define \cite{Abolfath,Piette} 
$\zeta_j = \varphi - \chi_j$ where $\varphi$ is the standard azimuthal
angle, e.g., ${\bf r}=(r\cos\varphi, r\sin\varphi)$
and $\chi_j$ is skyrmions orientation and measuring the deviation
from the standard hedge-hog fields. 
In Fig. \ref{Fig1}
the in-plane relative orientation, $\chi$, between two skyrmions 
with relative distance, {\bf R}, is shown.
If we have a gas of skyrmions far away from each other,
the total energy is invariant under variation of 
skyrmions orientation. For finite separation we expect a coupling  
between skyrmions due to different orientations.
Here we consider a situation 
where ${\bf m}_i$ are localized and well separated. 
This is valid for dilute skyrmions in a quantum Hall system, e.g., 
$\nu_{2c} < \nu < \nu_{1c}$. One may
divide the configuration space (${\rm R^2}$) into $N$ regions such
that $u_i$ is significant in region $i$ and small in others
\begin{equation}
u_{i\mu}=\frac{{\bf m}_{i\mu}}{1+\hat{z}\cdot{\bf m}_i}.  
\label{3}
\end{equation}
Here 
$\mu=(x,y)$ and 
\begin{equation}
{\bf m}_j=({\bf \varphi}_x^j, \varphi_y^j, 
\sqrt{1-\Phi_j\cdot\Phi_j}),
\label{4}
\end{equation}
where $j \neq i$ and
$\Phi_j=(\varphi_x^j, \varphi_y^j, 0)$
\begin{equation}
u_{j\mu}=\frac{\varphi_\mu^j}{1+\sqrt{1-\Phi_j\cdot\Phi_j}}.
\label{4.1}
\end{equation}
Substituting Eq.(\ref{2})-Eq.(\ref{4.1}) into Eq.(\ref{1}) and expanding
${\bf m}_j$ up to $\varphi_j$, we may find ${\bf m}$ in the $i$th region
\begin{equation}
{\bf m}={\bf m}_i + \Omega_i\times{\bf m}_i +
\frac{1}{2}\Omega_i\times(\Omega_i\times{\bf m}_i) +  
{\cal O}(\Omega^3),
\label{5}
\end{equation}
here Eq.(\ref{5}) describes an infinitesimal rotation of ${\bf m}_i$
about $\Omega_i$ axis where
\begin{equation}
\Omega_i=\frac{1}{2}{\bf m}_i\times\{
(1+\hat{z}\cdot{\bf m}_i) \Phi_{\rm eff}^i - 
({\bf m}_i\cdot\Phi_{\rm eff}^i)\hat{z}\},
\label{6}
\end{equation}
and 
\begin{equation}
\Phi_{\rm eff}^i({\bf r}) = \sum_{j \neq i}^N \Phi_j({\bf r}).
\label{7}
\end{equation}
Therefore the effect of the other skyrmions on the specific skyrmion
is the same as a single skyrmion 
 with charge $Q-1$ due to the effective
linear field, $\Phi_{\rm eff}$. 
In order to study the effect of the Zeeman and Coulomb energy,
we make the ansatz that the above superposition rule is valid 
for skyrmions far from each other even in the presence of 
the full interaction.
This can be taken into account by Eq.(\ref{5}) and evaluating
the total energy of multi-skyrmion in QHF.
It is obvious that the total energy can be divided into energies in 
separated regions. We assume that the interaction between the skyrmions
is weak, hence $\Omega_i={\bf m}_i \times \Phi_{\rm eff}^i$.
One may obtain easily the total energy by
redoing the same calculation for all separated regions                                                 
and sum over energies
\begin{eqnarray}
E[{\bf m}] &=& \sum_{i=1}^N \left(
E[{\bf m}_i] +
\int_i d^2{\bf r} \; \overline{\psi}_{\rm eff}^i \{ -\frac{\rho_s}{2} 
\nabla^2 + \frac{\tilde{g}}{4\pi\ell_0^2} \} \psi_{\rm eff}^i \right) 
\nonumber\\&& + E_{\rm eff}^C[{\bf m}],
\label{8}
\end{eqnarray}
where $\psi_{\rm eff}^i\equiv \Phi_{x \; \rm eff}^i + 
i \Phi_{y \; \rm eff}^i$ is a complex field
and $E_{\rm eff}^C[{\bf m}]$ is the effective Coulomb 
interaction between skyrmions. Note that in the absence of
the Coulomb term, the saddle point solution 
associated with the scalar field, $\psi^i$, are vortices, i.e.,
$-\nabla^2 \psi^i + \kappa^2 \psi^i=0$.
One may divide the total energy, Eq.(\ref{8}), into two parts,
the self energy of skyrmions and the interaction energy which are
designated by $T$ and $V$ respectively
\begin{equation}
T[{\bf m}]=\sum_{i=1}^N \left(
E[{\bf m}_i] +
\int_i d^2{\bf r} \sum_{j \neq i}^N
\overline{\psi}^j \{-\frac{\rho_s}{2} \nabla^2
+ \frac{\tilde{g}}{4\pi\ell_0^2} \} \psi^j \right).
\label{9}
\end{equation}
The first term in Eq.(\ref{9}) is the total self energy of the isolated 
skyrmions and 
the second term is the effect of their tail in other regions, 
i.e., the
contribution of their 
energy in regions far from the core.
Here we are interested to study the effect of 
interaction in a system of many skyrmions and their physical relevance.
One may find the effective interaction between the skyrmions by making
use of the Stokes theorem and neglecting 
the next nearest neighbor 
terms in Eq.(\ref{8})
which are described by the terms like
$\int_i d^2{\bf r} \sum_{j \neq i}\sum_{k \neq j} \Phi_j \cdot \Phi_k$
and $\int_i d^2{\bf r} \sum_{j \neq i}\sum_{k \neq j}
\partial_\alpha\Phi_j\cdot\partial_\alpha\Phi_k$ then
\begin{equation}
V_{\rm eff}[{\bf m}]= 
E_{\rm eff}^0[{\bf m}] + E_{\rm eff}^C[{\bf m}],
\label{11.0}
\end{equation}
and
\begin{eqnarray}
E_{\rm eff}^0[{\bf m}] &=& \sum_{i=1}^N
\int_i d^2{\bf r} \; \partial_a \{J_a^{\lambda(i)} \Omega_{i\lambda} \} 
\nonumber\\&&
= \frac{\rho_s}{2} \sum_{<ij>} \int_i d^2{\bf r} \overline{\psi}^j 
\{ -\nabla^2 + \kappa^2 \} \psi^i, 
\label{11}
\end{eqnarray}
where ${\bf J}^{(i)}$ has been defined for the $i$th skyrmion
and satisfies the Euler-Lagrange differential equations, Eq.(\ref{eq5})
and Eq.(\ref{eq6}).
$E^0_{\rm eff}[{\bf m}]$ is the contribution of the gradient and Zeeman energy
to the effective interaction. It describes a system of interacting dipoles
(see Fig. \ref{Fig3}).
Expanding the charge density of skyrmions, $\rho$, in terms of $\psi$,
leads to the effective Coulomb interaction

\begin{eqnarray}
E_{\rm eff}^C[{\bf m}] &=&  
\frac{e^2}{2\epsilon} \sum_{i \neq j}
\int_i d^2{\bf r} \int_j d^2{\bf r^\prime}
\frac{\rho_i({\bf r}) \rho_j({\bf r}^\prime)}{|{\bf r}-{\bf r}^\prime|}
\nonumber\\&&
+ \epsilon_{\mu\nu} \frac{\nu e^2}{4\pi\epsilon}
\sum_{i \neq j} \left\{\int_i d^2{\bf r} 
\partial_\mu V^j({\bf r})\partial_\nu{\bf m}_i \cdot \Omega_j \right.
\nonumber\\&& + \int_j d^2{\bf r} \left. 
\partial_\mu V^i({\bf r})\partial_\nu{\bf m}_j \cdot \Omega_i
+ \dots \right\},
\label{11.1}
\end{eqnarray}
where $\rho_i=\frac{-\nu}{8 \pi} \; \epsilon_{\alpha \beta} \; 
{\bf m}_i \cdot [\partial_\alpha{\bf m}_i \times 
\partial_\beta{\bf m}_i]$. 
The first and second term in Eq.(\ref{11.1}) are the 
Coulomb energy due to the
monopole and dipole counterparts of skyrmions respectively. 
The former 
falls off like $R^{-1}$ whereas the latter 
falls as $R^{-2}$ 
where the distance between two skyrmions is denoted by $R$.
Unlike the dense skyrmions in which the effect of the dipole term is crucial,
the monopole term dominates for an extremely dilute (i.e., $\nu \simeq 1$)
skyrmions system. Neglecting the Coulombic dipole term, 
one may evaluate the integrals in Eq.(\ref{11}) by 
applying the techniques that 
were developed for a pair of skyrmions
by Piette {\em et al.} \cite{Piette}
\begin{eqnarray}
V_{\rm eff}[{\bf m}] &=& 
\frac{e^2}{2\epsilon} \sum_{i \neq j}
\int_i d^2{\bf r} \int_j d^2{\bf r^\prime}
\frac{\rho_i({\bf r}) \rho_j({\bf r}^\prime)}{|{\bf r}-{\bf r}^\prime|}
\nonumber\\&& + \frac{c^2\tilde{g}}{4\pi^2} \sum_{<ij>}
\cos(\chi_j-\chi_i) K_0(\kappa |{\bf R}_j-{\bf R}_i|),
\label{12}
\end{eqnarray}
where $\chi_j-\chi_i$ and $R_j-R_i$ 
describe the in-plane relative orientation and effective distance 
between skyrmion the $j$th and the $i$th. The first term in Eq.(\ref{12}) is 
the electrostatic monopole term independent of the relative orientation.
The normalization factor, $c=30.4$, is calculated for a single skyrmion
numerically.
The second term in Eq.(\ref{12}) describes a classical XY-model 
where the minimum energy configuration specifies the 
relative orientation corresponds to $\chi_j-\chi_i=\pi$. 
$K_0(x)$ is the modified Bessel function, i.e., 
the coupling between the site $i$ and $j$ decays exponentially. 
We see that there is an exponential decrease of
the coupling between the $i$th and the $j$th skyrmions
for $R \gg 1/\kappa$.  
We will show that an antiferromagnet ordering of the hedge-hog fields
within centered square lattice is the minima
of the effective potential, Eq.(\ref{12}) \cite{Cote97}, in
some circumstances although the Coulomb interaction in Eq.(\ref{12})
is long range
and the XY term contributes to the total energy as a short range interaction.
The stability of the square lattice can be 
anticipated via generalizing the standard techniques for
calculating the collective modes which has been developed by the
Bonsall and Maradudine \cite{Bonsall} for the multi-skyrmion system.
\cite{Henk,Ramin}

\subsection{Recombination of the skyrmions}
\label{Recomb}

One may consider the static and dynamical properties of a skyrmion 
lattice using the interaction that we have derived in the
previous section
\begin{eqnarray}
V_{\rm eff} &=& \sum_{i \neq j}
  V_0(|{\bf R}_i-{\bf R}_j|) \nonumber\\&&
     + \sum_{<ij>} J(|{\bf R}_i-{\bf R}_j|) \cos(\chi_i - \chi_j).
\label{hh1}
\end{eqnarray}
In our model $V_0(R)$ is the electrostatic monopole interaction and 
$J(R) = \frac{c^2\tilde{g}}{4\pi^2} K_0(\kappa R)$. 
In Fig. \ref{Fig2}, $J$ is depicted
as a function of the Landau level filling factor for a given Zeeman energy. 
The sign of $V_{\rm eff}$ specifies the global minima of the energy
functional. For instance, if $V_{\rm eff}$ is positive then
$E(Q=N) > N E(Q=1)$ and the system can be composed by
seperated single skyrmions.
The validity of this inequality has been investigated recently by
Lillieh\"o\"ok {\em et al.}\cite{Lilliehook} when $N>2$. The case of
$N=2$ is marginal and will be discussed in the next subsection.
The {\em ionization energy} of the lattice per skyrmion
is identified by $E(Q=1)$ where $R \rightarrow \infty$.
Here $E(Q)$ is the optimized energy of an individual Q-skyrmion
(the self energy of the skyrmion).
One may immediately notice that 
the XY term in Eq.(\ref{hh1}) favors forming a Q-skyrmion ($Q=N$) by 
recombining the $N$ single skyrmions.
Note that the sign of this term in the antiferromagnet ordering is 
negative, therefore smaller seperation between skyrmions, $R$, is favorable.
Such a combination costs the Coulomb energy. 
Since the Coulomb interaction is long range
and the XY term contributes to the total energy as a short range interaction,
and the order of the Coulomb energy and the short range XY interaction
with respect to the number of particles are ${\cal O}(N^2)$ and 
${\cal O}(N)$ respectively, then
the stability of the lattice at the limit of large N as
a local minima is guaranteed
(the collapse of the lattice costs too much energy).
For large $N$, an antiferromagnet ordering
between the single skyrmions within a square lattice can be the
minima of the interaction energy.
Conversely, for the short range interactions, namely the hard-core model
\begin{equation}
E_c[{\bf m}] \propto \frac{1}{2} \int d^2 {\bf r}
\left( \partial_x {\bf m}({\bf r}) \times
\partial_y {\bf m}({\bf r}) \right)^2,
\end{equation}
we obtain $E(Q=N) < N E(Q=1)$, i.e., a charge-Q skyrmion is the
global minima, amongst all other configuarations. \cite{Piette} 
These models are convenient for the nuclear physics,
to describe the combination of the individual
nucleons to make a stable nuclei with the same baryonic number.
Here the baryonic number is analogues of the skyrmions charge.
This configuration which is favorable for the short range
strong interaction between the nucleons, can be destroyed by 
the long range Coulomb interaction.
Therefore the long range nature of the Coulomb interaction is crucial
for the stability of the lattice.
Varying the functional form of the interaction term, in $E_c$, 
changes the optimized value of the self energies for
different models of skyrmions and subsequently changes
the effective potential among them.
A cross over between a collapsed phase (no long range order phase)
into a Skyrme lattice occurs
due to changing the functional form of interaction, from short range
to a long range.
This cross over can be identified by 
the sign of the energy difference (the skyrmions self energy difference)
between the different topological spin configurations, $E(Q)-E(Q_2)-E(Q_1)$
where $Q=Q_1+Q_2$.
We end up this subsection with a comment. The results of our calculation show
$V_{\rm eff}$ is negligible as $\tilde{g}$ tends to zero.
For the zero Zeeman energy the Belavin-Polyakov solutions are the well
known analytical scale invariant solutions, i.e., any analytical function
possess the same energy and neither the skyrmion positions nor the
skyrmion size are given.\cite{Belavin,rajaraman,Nazarov}
In these circumstances, all of the possible $Q$-sector
configurations of a multi-skyrmion system
are degenerate, e.g., the energy of
the $N (=\sum_i Q_i)$ single skyrmions is equivalent to the charged-$N$
skyrmion. The seperated skyrmions do not feel
each other, hence there will be no effective potential among them.
$V_{\rm eff}$ is non-zero if $\tilde{g} \neq 0$.

\subsection{Triangular lattice of the bi-skyrmions}
\label{SubC}

An estimate based on the total energy of a pair of skyrmions with
the long range Coulomb interaction, shows that
the energy of a skyrmion with charge two is lower than the 
energy of two skyrmions each with charge one when
$\tilde{g} < \tilde{g}_c (=5.3 \times 10^{-6} e^2/\epsilon\ell_0)$.
\cite{Lilliehook}
For small values of the Zeeman energy
a pair of skyrmions is perefered.
Moreover, a bi-skyrmion
(charge-2 skyrmion) is stable to accretion of further charge.
One may expect a Skyrme lattice is governed via the formation
of the bi-skyrmions when $\tilde{g} < \tilde{g}_c$.
The inter-skyrmionic interaction between the bi-skyrmions
can be obtained by the same calculation which has been introduced
in Sec. \ref{SubA}. In this case, the relation between the hedge-hog
fields of the bi-skyrmions relative to the standard hedge-hog fields can be
defined by $\xi_j = 2\varphi - \chi_j$. 
Similar to the case of the single-skyrmion lattice,
we should have the correct shape of a charge-2 skyrmion to obtain
the effective potential.
We have solved Eq.(\ref{eq3}), using the finite-difference
method \cite{Abolfath} to find out the asymptotic behavior of a
circular symmetric bi-skyrmion. Far from its core we find
\begin{equation}
\theta(r) = \frac{c\kappa^2}{2\pi} K_2(\kappa r)
\end{equation}
where $K_2(x) = \sqrt{\pi/(2x)} e^{-x} (1 + {\cal O}(1/x))$ is
the modified bessel function. Our numerical calculation gives $c=79$.
It is straightforward to show that the interaction between the
bi-skyrmions can be obtained by:
\begin{eqnarray}
V_{\rm eff}[{\bf m}] &=& 
\frac{2 e^2}{\epsilon} \sum_{i \neq j}
\int_i d^2{\bf r} \int_j d^2{\bf r^\prime}
\frac{\rho_i({\bf r}) \rho_j({\bf r}^\prime)}{|{\bf r}-{\bf r}^\prime|}
\nonumber\\&& - \frac{c^2\tilde{g}^2}{8\pi^3} \sum_{<ij>}
\cos(\chi_j-\chi_i) K_0(\kappa |{\bf R}_j-{\bf R}_i|).
\label{pair}
\end{eqnarray}
We observe that the effective Coulomb interaction in the
multi bi-skyrmionic system
is a factor of $4$ of the single-skyrmionic lattice.
Moreover, the sign of the XY term is negative, rendering the possibly
ferromagnet state ($\chi_i = \chi_j$)
within the triangular lattice for small values of
the Zeeman splitting factor.
This result needs more comments. We have seen in Sec. \ref{SubA} that the
effective potential for the interaction between two seperated
single-skyrmions is similar to a pair of classical dipoles, i.e., a
single-skyrmion behaves as a source of the dipole fields.
On the other hand, the interaction form of a system of coupled
bi-skyrmions acts as a pair of quadrapole fields (a charge-2
skyrmion is a source of quadrapole fields). This is shown in Fig. \ref{Fig3}.
The minimum energy configuration is therefore ferromagnet alignment
of the quadrapoles.
This can explaine why the negative sign in the front of the bi-skyrmionic 
XY effective potential does exist.
However, the rest functional form of the interaction between the charge-$Q$
skyrmions is $Q$-independent. This stems from the behavior of the
asymptotic form of our linear field theory, \cite{Piette}
which has been implemented to derive the effective interaction among the
skyrmions. Note that the leading XY-term in \ref{pair} 
is the quadrapole-quadrapole
interaction, where the XY dipole-dipole term is absent.
This is originated from the non-trivial topological structure
of the charged-2 skyrmions.
The prediction of our classical field theory, therefore, rendering
a structural phase transition,
single-skyrmionic square lattice into bi-skyrmionic triangular lattice
at $\tilde{g}_c$.
We expect a first order phase transition occurs at $\tilde{g}_c$
where a latent heat is needed.
However, one may expect this phase transition can be affected by other
mechanisms, like skyrmion scattering from the impurities
and the quantum disordering.
Since the energy differences needed for observing
this structural phase transition
are very small, due to extremely small Zeeman energy,
then one expects the quantum fluctuations which are present at zero
temperature destroys the long range-order of the bi-skyrmion lattice.
This requires further investigations, one should
include such processes to the present formalism in order to obtain a
realistic phase diagram. 
The results of our calculation for the stability of the bi-skyrmion
lattice against the quantum fluctuations is still on the way and will
be presented elsewhere. \cite{Ramin}
Note that the sign of the XY-term in Eq.(\ref{pair}) in the ferromagnet
ordering is negative. The collapsed N-charged skyrmion is therefore prefered
for the short range interaction models, like the hard-core model.

\subsection{Triangular lattice of the single-skyrmions}

Obviously, the skyrmion density may be controlled 
via the Landau level filling factor.
For larger separations, the XY term in Eq.(\ref{12}) is
therefore negligible compared to the effective Coulomb energy. 
This implies that for a lower density of the single skyrmions  
the Coulomb interaction determines the lattice
structure and leading to a triangular lattice, assuming
$\tilde{g} > \tilde{g}_c$.
However, for a more dense system one may expect a 
different stable structural symmetry, as we have mentioned before.
Unlike the case where the 
competition between the monopole Coulomb term and the 
gradient dipole terms is crucial 
to the stability of the square lattice,
the electrostatic monopole interaction is the dominant 
term to determine the structure of the lattice for an 
extremely dilute ($\nu \sim 1$) skyrmions,
where the Coulomb interaction is long range and the XY term contributes
to the total energy as a short range interaction.
In this situation a specific 
skyrmion moves in a background which 
is being made by the vacuum of the other
skyrmions. Since $u_i$ which has been defined by Eq.(\ref{3}),
is zero everywhere but in the $i$th region,
then $\rho({\bf r})=\sum_i \rho_i({\bf r})$. 
Clearly the total charge is the summation upon the  
individual skyrmion's 
charge, i.e., 
$Q_{\rm tot}=\sum_i Q_i$ where $Q_i$ is the topological 
charge of localized skyrmion in that region. In this case,
the interaction between separated skyrmions is independent
of the relative orientation.
The skyrmions may be considered by point particles where
$\rho_i({\bf r})=\nu \delta({\bf r}-{\bf R}_i)$ and
\begin{eqnarray}
V_{\rm eff} &=&
\frac{e^2}{2\epsilon} \sum_{i\neq j}^N
\int d^2{\bf r}\int d^2{\bf r}^\prime \;
\frac{\rho_i({\bf r})\rho_j({\bf r^\prime})}
{|{\bf r}-{\bf r}^\prime|} \nonumber\\&&
=\frac{\nu^2 e^2}{2\epsilon} \sum_{i\neq j}^N
\frac{1}{|{\bf R}_i-{\bf R}_j|}.
\label{15}
\end{eqnarray}
Therefore the ground state is being described by Eq.(\ref{15}) is clearly
a triangular lattice, independent of the skyrmion orientation, 
i.e., a  Wigner crystal. \cite{Bonsall} This is in agreement with the
field theoretic result of Green {\em et al.} \cite{Green} and
the microscopic pictures. \cite{Brey,Cote97}
One may find the critical filling factor $\nu_{1c}$ 
where a transition into square lattice takes place.
We define a domain of separation 
for which the 
square lattice is valid, e.g.,
$\nu_{2c} < \nu < \nu_{1c}$ or $\kappa^{-1} < R < R_0$. 
Here the length scale 
cut off is denoted by $R_0$, beyond that a phase transition
to a triangular lattice occurs. 
Note that we have made an approximation 
associated with $R > \kappa^{-1}$ to demonstrate the 
existence of the square lattice. 
We may expect the approximation, namely the linear
superposition rule, Eq.(\ref{5}), fails
for a high density skyrmions. The crossover between low 
and high density of skyrmions 
occurs at $\kappa R \sim 1$
and leading to the lower critical filling factor $\nu_{2c}$, i.e.,
$\nu_{2c} = 1 - \tilde{g}/2\rho_s$.
A comparison between the ground state energy per skyrmion for the
square lattice and for the triangular lattice 
yields an estimate for the critical filling factor, $\nu_{1c}$,
the square-into-triangular transition point at $\tilde{g}(>\tilde{g}_c)$.
The result obtained from this calculation leads to a condition for
the stability of the square lattice
\begin{equation}
\tilde{g}_0^{3/4} |1-\nu|^{-1/4}
\exp\left\{-6.333\sqrt{\frac{\tilde{g}_0}{|1-\nu|}}\right\}
\geq 2 \times 10^{-4}.
\end{equation}
Here $\tilde{g}_0 = \tilde{g}/(e^2/\epsilon\ell_0)$.
Using the typical Zeeman energy,
$\tilde{g} = 0.015 e^2/\epsilon\ell_0$ for GaAs and taking advantage 
of the skyrmion-antiskyrmion duality, we obtain $|\nu_{1c}-1| \sim 0.02$
and $|\nu_{2c}-1| \sim 0.3$. 
Our estimate for the structural phase transition at $\nu_{1c}$ 
is a factor of 10 smaller than the microscopic Hartree-Fock result
\cite{Cote97}, although
it shows qualitative agreement with that microscopic picture.

\subsection{Quantum phases}

The appropriate quantum mechanical spectrum of the Skyrme lattice 
in our model can be obtained 
through Eq.(\ref{12}) and the Wess-Zumino action
\begin{eqnarray}
&&\left\{ \sum_{i=1}^N
\frac{\hbar^2}{2}\Lambda_0^{-1}
\left(-i \frac{\partial}{\partial \chi_i} - \xi_0 \right)^2
+ \sum_{i \neq j} V_0(|{\bf R}_i-{\bf R}_j|)
\right.\nonumber\\&& \left.
+ \sum_{<ij>} J(|{\bf R}_i-{\bf R}_j|)\cos(\chi_i-\chi_j) \right\} 
\Psi([{\bf R}; \chi]; \tau)
\nonumber\\&&  
= - \frac{\hbar}{8\pi} \frac{\partial}{\partial\tau}
\Psi([{\bf R}; \chi]; \tau),
\label{12.4a}
\end{eqnarray}
where  $[{\bf R}; \chi] \equiv 
({\bf R}_1; \chi_1, {\bf R}_2; \chi_2, ..., {\bf R}_N; \chi_N)$
and ${\bf R}_i \equiv (R_i\cos\varphi_i, R_i\sin\varphi_i)$.
$\varphi_i$ is the standard azimuthal angle, indicating 
the $i$th skyrmion. $\Lambda_0=\hbar^2\xi_0/(3\tilde{g})$ is the moment of
inertia of a single skyrmion, and $\xi_0 = E_z/(2\tilde{g})$. 
Obviously, Eq.(\ref{12.4a}) describes a quantum rotor problem for 
$\chi$-degree of freedom.\cite{Steve} This equation which has been obtained 
from a {\em first-principle} calculation is consistent with the result
of Ref. 27, and 28 which has been estimated phenomenologically via   
the fitting the known Hartree-Fock long wave length
spin waves spectrum.
The ground state described by Eq.(\ref{12.4a}) 
for skyrmions orientation is a quantum antiferromagnet where 
the U(1) symmetry is broken spontaneously.
This model has been used to describe the superconductor-insulator (SCI) 
quantum phase transition in granular
superconductors and Josephson junction arrays, \cite{Cha}
and predicts presumably
a continuous quantum phase transition in QHF.\cite{RMP}
The appropriate parameters which control the SCI phase transition
at zero temperature is the Zeeman splitting ($\tilde{g}$) and the
Landau level filling factor.
Using the empirical 
data for practical systems
($\tilde{g}=0.015 e^2/\epsilon\ell_0 \sim 2 K$)
shows the XY coupling constant, 
$J = (c^2\tilde{g}/4\pi^2) K_0(\sqrt{\tilde{g}/2\pi\ell_0^2\rho_s} \; R)$,
is a fraction of the $U = (\hbar^2/2\Lambda_0) \sim 0.4 K$ within
$\nu \gtrsim \nu_{1c}$. It
leads to the prediction of the Mott-insulating phase for this system
where we assume that the position of skyrmions are fixed,
e.g., they are crystalized.
Here $E_z = 0.214 e^2/\epsilon\ell_0$ is the Zeeman energy associated with
the optimal single skyrmion solution corresponding to 
$\tilde{g}=0.015 e^2/\epsilon\ell_0$, is evaluated by solving
Eq.(\ref{eq3}), for a single skyrmion. 
The details of this calculation have been
presented elsewhere. \cite{Abolfath}
We also expect that the system go through superconducting phase
for $J$ greater than $U$ since the skyrmions carring
the electrical charges.
In Fig. \ref{Fig2}, $J$ and $U$ are depicted
in terms of the Landau level filling factor for a given Zeeman energy. 
One may see that 
at $\nu=1$ the XY-interaction between the skyrmions is negligible.
As it is shown in Fig. \ref{Fig2} there are 
two symmetric critical point around $\nu=1$.
This symmetry is due to the duality of the quantum Hall 
skyrmion-antiskyrmion at $\nu=1$.
The nature of this phase transition could be
anticiapted via the bosonic Hubbard model of the quantum rotors. \cite{Cha}
Its critical behavior can be charecterized via a map onto 
a $2+1$D classical XY model. \cite{RMP}
Note that $\xi_0$ is a continuous variable. It changes continuously
with the size of skyrmions and/or the Zeeman energy.
For the half integer value of $\xi_0$ where $2\xi_0$ is odd,
our model leads to the Mott-insulating 
phase about $\nu=1$ 
for the quantum Hall liquid where the classical ordered phase is destroyed
by the quantum fluctuations.
Within the quantum rotors model, the skyrmions are analogous to the vortices.
This is a dual picture for the quantum Hall state, where the Mott-insulating
gap of the bosons and the super-current of the vortices are equivalent to
the quantum Hall gap and the quantum Hall current, repectively
\cite{Abolfath1},
consistent with the microscopic picture of Ref. 27. 
The energy gap of
the Mott-insulating phase, $\Delta \equiv 2U (= 6\tilde{g}^2/E_z)$,
is the analogous of the long wave length (${\bf k} \rightarrow 0$)
energy gap of the magnons. Magnons are the appropriate
collective modes of the quantum Hall liquid corresponding
to a single flipped spin, propagating throught the system.
The numerical study for the single skyrmions 
\cite{Abolfath} shows that $E_z \propto (e^2/\epsilon\ell_0)^{2/3} 
\tilde{g}^{1/3}$ as $\tilde{g} \rightarrow 0$. This leads to the
prediction of vanishing of the Mott-insulating gap as 
$\Delta \rightarrow \tilde{g}^{5/3}$.
We estimate $\Delta \sim 1K$ at $\tilde{g} = 0.015 e^2/\epsilon\ell_0$.
Therefore for the zero Zeeman splitting ($U=J=0$), 
the effective Hamiltonian is reduced to the exchange Coulomb 
interaction among the electrons. In this case,
the gapless Goldstone modes corresponding to the spontaneous 
global $SU(2)$ symmetry breaking (spin waves) are responsible for the 
incompressibility of the quantum Hall liquid.
The stability of the quantum Hall state can be investigated via
evaluating the dispersion curve of the magnons.

\subsection{Melting of the Skyrme lattice}
\label{Sec_Me}
In the following subsection, we implement the theory of elasticity
\cite{Landau} to study the non-zero temperature phase transitions, and
the zero-temperature quantum fluctuations.
The universality class of the phases associated with the 
Hamiltonian, Eq.(\ref{12.4a}),
can be changed when $T > 0$. The Skyrme lattice can be anticipated
via the long-range order of the position of the skyrmions 
as well as the orientational long-range order of their hedge-hog fields.
These long-range orders can be destroyed at finite temperature.
One may show,\cite{Ramin} the fluctuations of the skyrmion
orientation fields diverge logarithmically,
$\langle \chi^2 \rangle \sim T \ln(L/R)$, where
$L$ is the system size. This characterize the quasi-long range 
order of the skyrmions orientation fields on the $2D$-plane.
Forming a pair of the vortex-antivortex
from the hedge-hog fields, leads to a finite
temperature Kosterlitz-Thouless phase transition ($T_{KT}$),
assuming that skyrmions are crystalized. \cite{Cote97}
We may predict a critical temperature
for the KT phase transition based on the classical $2$D XY-model.
Our calculation for the classical 2D nearest neighbor XY-model 
on the square lattice, shows that the phase transition occurs
around $T_c = 0.9 J \sim 4 K$ at $\nu=0.8 \; (B_0 \sim 7 T)$.
One may also expect the 2D-Skyrme lattice melts (at $T=T_m$) via the
Kosterlitz-Thouless-Halperin-Nelson-Young (KTHNY) mechanism.\cite{Green}
In this case the disorder phase can be emerged via mediating a pair 
of dislocations. 
Similarly, the fluctuations of displacement
fields diverge logarithmically,\cite{Ramin}
$\langle R^2 \rangle \sim T \ln(L/R)$.
This exhibits a finite temperature quasi-long range positional order.
Our classical field theoretic results for the $T_{KT}$ and $T_m$
are shown in Fig. \ref{Fig4}.
The details of these calculations
will be presented elsewhere.\cite{Ramin}
The numerical calculation shows that the KT phase
transition can not account for this system except for small $|1-\nu|$
($\nu \gtrsim \nu_{1c}$) and/or large $\tilde{g}$ where $J < k_B T_m$,
as depicted in Fig. \ref{Fig4}.
In contrast to the case of small Zeeman energy where the KTHNY is 
responsible for destroying the translational long range order
(and therefore melting the Skyrme lattice),
the orientational long-range order of the hedge-hog fields
can be destroyed through the KT transition when the Zeeman energy
is large.
However, more accurate calculation is needed to obtain
the melting temperature of the Skyrme lattice. 
For this, we perform a Monte Carlo simulation. 
The results will be presented elsewhere.
By taking the zero point quantum fluctuations into account, the Hamiltonian
Eq.(\ref{12.4a}) leads to $\langle R^2 \rangle \sim |1-\nu| a^2 /5$ at $T=0$,
where $a$ is the lattice spacing. Using the Lindemann criterion,
we observe that the zero point quantum fluctuations destroys the long range
order of the Skyrme lattice when $|1-\nu| \geq 0.5$. This is the high
density limit of the skyrmions where our linear field theory may fail.

\section{Conclusion} 
In this paper we presented the analytical form of the 
inter-skyrmionic interaction via a {\em first-principle} calculation
within the classical field theory. The effective Hamiltonian 
demonstrated that for extremely dilute skyrmions, the ground state
is a triangular lattice to minimize the Coulomb repulsion. However
at higher densities, a square lattice forms to optimize the spin 
gradient and Zeeman energies. 
This work has provided a possibility 
that a triangular to a square lattice phase transition
which has been previously derived, using the Hartree-Fock
method, can also be revealed by the field theory picture.
In contrast to Ref. 11, 
we show that the field theoretic description of the NL$\sigma$-model 
does indeed lead to quantitative conclusion in 
agreement with microscopic Hartree-Fock calculations.
We also argued the possibility of the 
superconducting-insulator phase transition.
We have shown that within some special short range interactions the 
level-crossing between different kind of skyrmions may occur 
to destablize the Skyrme lattice, and
a single-skyrmion with the same topological charge can be formed.
We have shown that for extremely small Zeeman energy, skyrmions bound in
pair to make a triangular lattice with the ferromagnets orientation.
The possibilty of the Kosterlitz-Thouless phase transition as well as
the Kosterlitz-Thouless-Halperin-Nelson-Young phase transition has been
investigated.

\section{Acknowledgement}
Our special thanks go to S.M. Girvin and H.T.C. Stoof for
inspiring comments and ideas on the subject and for providing copies
of their unpublished works.
The authors thank H.A. Fertig, A.H. MacDonald, J.J. Palacios, 
and S. Rouhani for helpful discussion.
MA would like to thank Indiana University for the warm hospitality.

\begin{figure}
\caption{
The in-plane relative orientation, $\chi$, between two skyrmions 
with relative distance, {\bf R}, is shown.
}
\label{Fig1}
\end{figure}
\begin{figure}
\caption{
Minima configuration of the system of a coupled single- (a) and
bi-skyrmions (b) are shown. The single (bi) skyrmion act as a
source of the classical dipole (quadrapole) fields, far from its
core.
}
\label{Fig3}
\end{figure}
\begin{figure}
\caption{
The hopping term of the Bosonic Hubbard model, $J$, and 
the charging energy $U$ as a function of the
Landau level filling factor are depicted for a typical
Zeeman splitting factor.
}
\label{Fig2}
\end{figure}
\begin{figure}
\caption{
The prediction of the classical field theory for
the disordering temperature as a function of the Landau
level filling factor is shown for $\tilde{g} = 0.015 e^2/\epsilon\ell_0$
and $\tilde{g} = 0.1 e^2/\epsilon\ell_0$ (the inset). 
In contrast to the case of small Zeeman energy where the mechanism of the
KTHNY is responsible for the melting of the Skyrme lattice,
the orientaional long-range order of the hedge-hog fields can be
destroyed through the KT transition for the large Zeeman energies.  
}
\label{Fig4}
\end{figure}

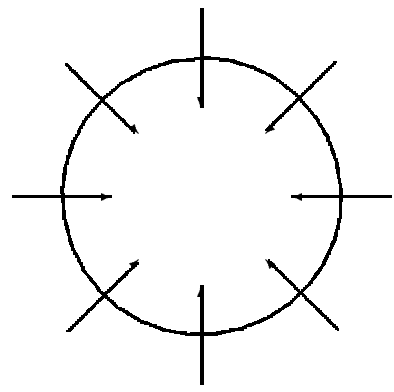
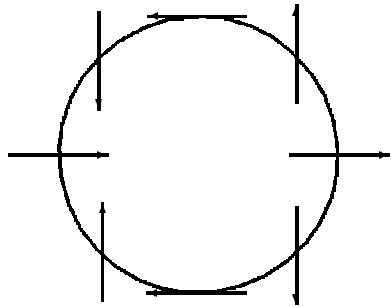
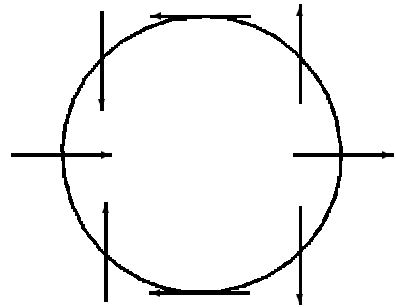
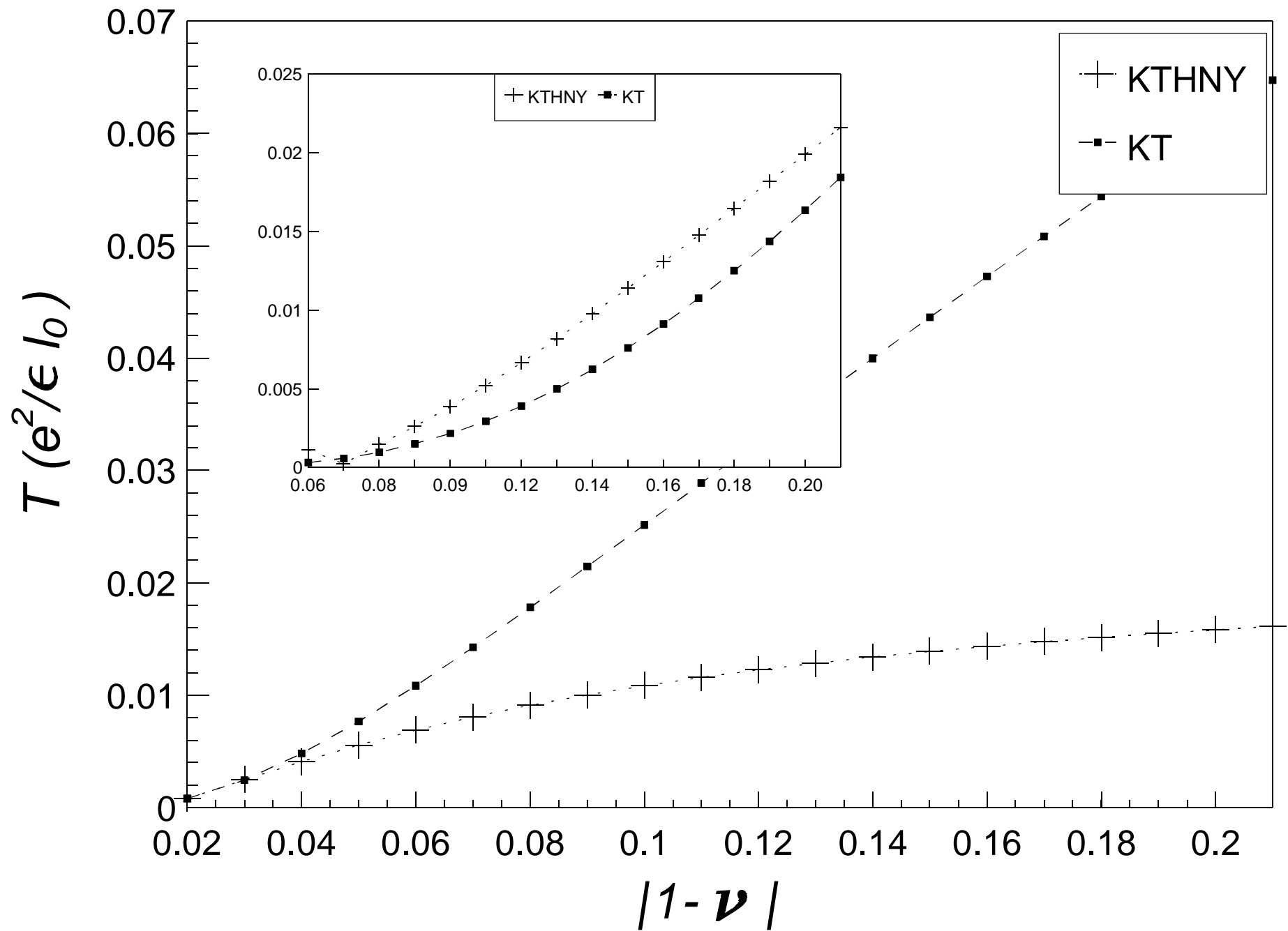
\end{document}